\documentclass[10pt,aps,pra,longbibliography,twocolumn]{revtex4-2}
\usepackage{longtable}
\usepackage{dcolumn}
\usepackage{times}
\usepackage{nicefrac}
\usepackage{amsmath}
\usepackage{amsfonts}
\usepackage{amssymb}
\usepackage{amsthm}
\usepackage{bm}
\usepackage{bbm}
\usepackage{physics}
\usepackage{graphicx} 
\usepackage[table,xcdraw,dvipsnames]{xcolor}

\newcolumntype{.}{D{x}{}{-1}}
\newcolumntype{w}[1]{D{.}{.}{#1}}

\newcommand{\Za}{Z\alpha}

\newcommand{\vare}{\varepsilon}

\newcommand{\pr}{^{\prime}}

\newcommand{\bfp}{{\bm p}}
\newcommand{\bfq}{{\bm q}}
\newcommand{\bfr}{{\bm r}}

\newcommand{\bfpp}{{\bm p}\pr}

\newcommand{\balpha}{\bm{\alpha}}

\newcommand{\bnabla}{\bm{\nabla}}

\newcommand{\hp}{\hat{\bfp}}

\newcommand{\hpp}{\hat{\bm p}\pr}

\newcommand{\SixJ}[6]{
        \left\{
        \begin{array}{ccc}
        #1  & #2  & #3 \\
        #4  & #5  & #6 \\
        \end{array}
        \right\}
        }

\begin{document}
\title{Self-energy correction to the $\bm{E1}$ transition amplitudes in hydrogen-like ions}
\author{M. G. Kozlov$^{1, 2}$}
\author{M. Y. Kaygorodov$^1$}
\author{Yu. A. Demidov$^{1, 2}$}
\author{V. A. Yerokhin$^3$}

\affiliation{
$^1$ Petersburg Nuclear Physics Institute NRC ``Kurchatov Institute'', Gatchina 188300, Russia \\
$^2$ St.~Petersburg Electrotechnical University ``LETI'', St. Petersburg 197376, Russia\\
$^3$ Peter the Great St. Petersburg Polytechnic University, St. Petersburg 195251, Russia
}
\begin{abstract}
We present calculations of the self-energy correction to the $E1$ transition amplitudes in hydrogen-like ions, performed to all orders in the nuclear binding strength parameter. Our results for the $1s$-$2p_{1/2}$ transition for the hydrogen isoelectronic sequence show that the perturbed-orbital part of the self-energy correction provides the dominant contribution, accounting for approximately 99\% of the total correction for this transition. Detailed calculations were performed for $ns$-$n'p$ and $np$-$n'd$ transitions in H-like cesium. We conclude that the perturbed-orbital part remains dominant also for other $ns$-$n'p$ transitions, whereas for the $np$-$n'd$ matrix elements this dominance no longer holds. Consequently, the self-energy corrections for the $np$-$n'd$ one-electron matrix elements cannot be well reproduced by means of effective QED operators constructed for energy levels. 
\end{abstract}
\maketitle

\section{Introduction}

The rapid progress of experimental techniques in atomic spectroscopy, combined with improved computational methods developed for many-electron systems, requires the inclusion quantum electrodynamics (QED) corrections, not only for energy levels but also for other properties of many-electron atoms. In particular, recent large-scale many-body calculations of $E1$ transition amplitudes for neon-like iron and nickel have reached such high computational precision that the inclusion of QED effects has become mandatory \cite{Kuhn2020,KCO22,SKBS24}.

An accurate knowledge of the transition amplitudes in many-electron atoms is necessary for the atomic tests of the standard model and in the search for the new physics beyond standard model \cite{SBDJ18}. For example, it is important to include QED corrections to the parity non-conserving transition amplitude in Cs \cite{STP05} when comparing theory with the experiment \cite{WBC97}. QED corrections to the dynamic polarizability may affect magic frequencies for the clock transitions.

So far ab initio QED calculations have been conducted primarily for atoms and ions with one or several electrons; extending such methods to many-electron systems looks unfeasible at present. It is possible to perform ab initio QED calculations in many-electron systems  within one-electron approximation \cite{sapirstein:02,sapirstein:03:hfs}, but the omitted electron-correlation effects can be very significant in this case.

Most of recent calculations of $E1$ transition amplitudes of many-electron atoms performed so far \cite{KCO22,SKBS24,Fairhall2023,Roberts2023,TaDe23,KoDeKa24} accounted for QED effects by using approximate methods based on different variants of effective QED potentials. The most successful of these is the model operator introduced by Shabaev and coworkers in Ref.~\cite{STY13} and implemented as QEDMOD package \cite{STY15,shabaev:18:qedmod}. It is important that the QEDMOD operator, as well as all other effective QED potentials, is designed to approximately reproduce QED effects for energy levels only. As a result, using these operators for calculations of  transition amplitudes yields incomplete results, as they account only for part of the effect while neglecting other contributions, such as vertex and reducible corrections.

To justify the omission vertex and reducible corrections, the studies \cite{KCO22,SKBS24,Fairhall2023,Roberts2023} cited the work of Sapirstein and Cheng \cite{SaChe05}, who performed ab initio calculations of $E1$ transition amplitudes for alkali atoms within the one-electron screening-potential approximation and found these corrections to be relatively small. However, Sapirstein and Cheng’s study focused solely on the $ns$-$np_{1/2}$ transition and was performed for a few alkali-like atoms, leaving the broader applicability of this statement uncertain. In our previous work \cite{KoYeKa24}, we performed one-electron ab initio calculations for neon-like iron and similarly concluded that the vertex and reducible corrections are small enough to be neglected at the current level of precision. Nonetheless, the general validity of this conclusion remains unclear.

The QED correction to the $E1$ transition amplitude consists of the electron self-energy and the vacuum-polarization parts. The vacuum-polarization part is relatively simple and can be easily accounted for, e.g., by means of vacuum-polarization potentials included into the QEDMOD package. By contrast, the self-energy part is much more complicated as it cannot be successfully approximated by a local potential.

In this work, we conduct a detailed study of the self-energy correction to the $E1$ transition amplitudes in hydrogen-like ions, with a particular focus on hydrogen-like cesium. The total self-energy contribution is divided into two parts: the perturbed-orbital (po) part, which can be approximately represented by effective QED potentials, and the vertex+reducible (vr) part, which is omitted by them. We analyze the relative effect of the vr part on the total self-energy correction for various transitions. Additionally, we examine the accuracy of the QEDMOD potential in approximating ab initio self-energy results. This information is required for assessing potential errors when using the QEDMOD operator in many-electron calculations of the $E1$ transition amplitude. 

In this paper we use the relativistic units $m=\hbar = c = 1$ and the Heaviside charge units $\alpha = e^2/(4\pi)$.

\section{Basic formulas}

The $E1$ transition amplitude in the length gauge between the one-electron states $a$ and $b$ is given by the expectation value of the electric dipole operator,
\begin{align}\label{eq:zab}
z_{ab} = \bra{a} r_z \ket{b}\,,
\end{align}
where $r_z$ is the $z$ component of the position vector $\bm{r}$.

The self-energy correction to the matrix element $z_{ab}$ is 
represented as a sum of the perturbed-orbital (po) and the vertex$+$reducible (vr) parts,
\begin{align}
\delta z_{\rm se} = z_{\rm po} + z_{\rm vr}\,.
\end{align}
General formulas for these corrections were reported in the literature \cite{SaChe05} (see also Ref.~\cite{2002_Shabaev}).
The perturbed-orbital part is given by
\begin{align}
z_{\rm po} = &\
    \sum_{n\neq a}\bra{a} \Sigma_R(\vare_a) \ket{n}\,
        \frac{z_{nb}}{\vare_a-\vare_n}
    \nonumber \\ &
+    \sum_{n\neq b}
        \frac{z_{an}}{\vare_b-\vare_n}
        \bra{n} \Sigma_R(\vare_b) \ket{b}\,,
\end{align}
where $\Sigma_R(\vare)$ is the renormalized one-loop self-energy operator, $\Sigma_R(\vare) = \Sigma(\vare)-\gamma^0\delta m$,
and $\delta m$ is the one-loop mass counterterm.
The one-loop self-energy operator is defined by its matrix elements with one-electron wave functions
as \cite{1999_Yerokhin}
\begin{eqnarray} \label{se6}
\bra{a} \Sigma(\vare) \ket{b} = \frac{i}{2\pi} \int_{-\infty}^{\infty}d\omega\,
 \sum_n \frac{\bra{an} I(\omega) \ket{nb}}{\vare-\omega -u\vare_n}\,,
\end{eqnarray}
where $u = 1-i0$, the summation is performed over the complete spectrum of the Dirac equation and $I(\omega)$ describes the
exchange by a virtual photon. In the Feynman gauge, $I(\omega)$ is given by
\begin{eqnarray} \label{se6b}
I(\omega,\bfr_1,\bfr_2) = \alpha \big( 1-\balpha_1\cdot \balpha_2\big)\, \frac{e^{i\sqrt{\omega^2+i0}\,r_{12}}}{r_{12}}\,,
\end{eqnarray}
with $r_{12} = |\bfr_1-\bfr_2|$.

The vertex$+$reducible contribution is given by
\begin{align}\label{eq6}
z_{\rm vr} = &\
    \frac{i}{2\pi}\int_{-\infty}^{\infty}d\omega\,
  \nonumber \\ &
    \times
    \Bigg[
        \sum_{n_1n_2}\frac{z_{n_1n_2}\, \bra{an_2} I(\omega) \ket{n_1b}}
            {(\vare_a-\omega-u\vare_{n_1})(\vare_b-\omega-u\vare_{n_2})}
  \nonumber \\ &
  - \frac{z_{ab}}{2}\,\sum_{n}\frac{\bra{an} I(\omega) \ket{na}}
            {(\vare_a-\omega-u\vare_{n})^2}
  \nonumber \\ &
  - \frac{z_{ab}}{2}\,\sum_{n}\frac{\bra{bn} I(\omega) \ket{nb}}
            {(\vare_b-\omega-u\vare_{n})^2}
  \Bigg]\,.
\end{align}

\subsection{Angular integration}

Using the standard Racah algebra, we evaluate the leading-order $E1$ transition matrix element
$z_{ab}$ as
\begin{align}\label{ang:1}
z_{ab} = &\ \frac{(-1)^{j_b-\mu_b}}{\sqrt{3}}\,
        C_{j_a,\mu_a,\, j_b,-\mu_b}^{1,0}\,
        C_1(\kappa_a,\kappa_b)\,r_{ab}
\nonumber \\
  \equiv &\ {\cal P}_{ab}\,r_{ab}\,,
\end{align}
where $C_{j_1,m_1,\, j_2,m_2}^{j,m}$ is the Clebsch-Gordan coefficient, $C_L(\kappa_1,\kappa_2)$
is the reduced matrix element of the normalized spherical harmonics, see, e.g., Eq.~(A10)
of Ref.~\cite{yerokhin:20:green}, and the radial integral is
defined as
\begin{align}\label{eq:rab}
r_{ab} =  \int_0^{\infty}dr\, r^3\,\big[ g_a(r)\,g_b(r)+f_a(r)\,f_b(r)\big]\,,
\end{align}
where $g(r)$ and $f(r)$ are the upper and lower radial components of the Dirac wave functions.
The angular factor ${\cal P}_{ab}$ introduced in Eq.~(\ref{ang:1}) is the common prefactor for all corrections
to $z_{ab}$.

For the evaluation of the vertex correction, we need to perform the angular integration in the following expression
\begin{align}
X_{n_1n_2} = \sum_{\mu_1\mu_2} z_{n_1n_2}\, \bra{an_2} I(\omega) \ket{n_1b}\,,
\end{align}
where $\mu_1$ and $\mu_2$ are the momentum projections of the states $n_1$ and $n_2$, respectively.
We make use of the standard partial-wave decomposition of the matrix element with $I(\omega)$
\begin{align}
\bra{an_2} I(\omega) \ket{n_1b} = \alpha \sum_L
J_L(an_2n_1b)\, R_L(\omega,an_2n_1b)\,,
\end{align}
where $J_L$ contains all the dependence on the angular momenta projections and given by Eq.~(39)
of Ref.~\cite{yerokhin:20:green} and $R_L$ are the standard two-body radial integrals
defined in Appendix A of Ref.~\cite{yerokhin:20:green}. After performing summations over the
momentum projections with help of formulas from \cite{varshalovich}, we obtain
\begin{align}
X_{n_1n_2} = &\ \alpha\, {\cal P}_{ab}\, \sum_L r_{n_1n_2}\,R_L(\omega,an_2n_1b)\,
  \nonumber \\ &
    \times
    \frac{C_1(\kappa_1,\kappa_2)}{C_1(\kappa_a,\kappa_b)}\,
    (-1)^{j_1-j_2}\,
    \SixJ{j_2}{j_1}{1}{j_a}{j_b}{L}\,.
\end{align}

\subsection{Momentum-space reduction}

In order to perform renormalization in the vertex$+$reducible term, we need to separate out the contribution
of the free electron propagators and evaluate them in momentum space, with a covariant regularization of
ultraviolet divergences. The divergent terms can be shown to cancel between the vertex and reducible parts,
whereas the remaining finite contribution needs to be calculated in momentum space.
So, we represent  Eq.~(\ref{eq6}) as a sum the free part that contains the contribution of free electron propagators and the many-potential part which is the remainder,
\begin{align}
z_{\rm vr}
=
z_{\rm vr}^{(0)}
+ z_{\rm vr}^{(1+)}. 
\end{align}

The general calculation procedure is similar to that developed in the previous studies, see, e.g., Ref.~\cite{yerokhin:99:sescr}. New features appear only in calculations of the free vertex part, which contains the vertex operator with the electric dipole operator in the momentum space. We therefore will describe this part of our calculation in some detail. 

The Fourier transform of the electric dipole operator is given by the gradient of the Dirac $\delta$ function,
\begin{align}
{\bm r}_z \stackrel{\rm Fourier}{\longrightarrow}  i(2\pi)^3\,\bnabla_z\,\delta^3(\bfq)\,,
\end{align}
where $\bfq = \bfp-\bfpp$ is the exchanged momentum.
It is rather cumbersome to perform calculations with the above momentum-space expression for the electric dipole operator, because we need to perform the integration by parts analytically before carrying out the integration over the momentum in the vertex operator.
For this reason, we choose to follow Ref.~\cite{persson:97:g} and introduce a finite regularization parameter $\rho$, which makes the Fourier transform to be a continuous function,
\begin{align}
{\bm r}_z\,e^{-(\nicefrac{\rho\,r}{2})^2} \stackrel{\rm Fourier}{\longrightarrow}
   -i\,\bfq_z\,\frac{16\pi^{3/2}}{\rho^5}\,e^{-\nicefrac{q^2}{\rho^2}}\,.
\end{align}
We found that if we perform numerical calculation with a sufficiently small regularization parameter
(typically, $\rho \approx 10^{-6}$), the error introduced by the regulator is completely negligible
at the level of our interest. Moreover, only a small momenta region $|p-p'| \sim \rho$ contributes to the total integrals, so that numerical integrations are relatively simple in this case.

We thus write the free part of the vertex contribution as
\begin{multline}
z_{\rm ver}^{(0)} =  -i\, \frac{16\pi^{3/2}}{\rho^5}\,
 \int \frac{d^3\bfp}{(2\pi)^3}
 \int \frac{d^3\bfpp}{(2\pi)^3}\,
 e^{-\nicefrac{q^2}{\rho^2}}\,
\\ 
\times
 \overline{\psi}_a(\bfp)\,
 \Gamma_R^0(\vare_a,\bfp;\vare_b,\bfpp)\,
 \big(\bfp_z-\bfpp_z\big)\,
 \psi_b(\bfpp)\,,
\end{multline}
where $\overline{\psi}_a(\bfp) = \psi^{\dag}_a(\bfp)\gamma^0$ and $\Gamma_R^0$ is the time component of the renormalized
free-electron vertex operator, see Appendix B of Ref.~\cite{yerokhin:99:pra} for definition and explicit representation.
We now use the following representation for the vertex operator sandwiched between two Dirac functions,
\begin{multline}
\overline{\psi}_a(\bfp)\, 
 \Gamma_R^0(\vare_a,\bfp;\vare_b,\bfpp)\,
 \psi_b(\bfpp) =
 \frac{\alpha}{4\pi}\,i^{l_a-l_b}\,
\\ \times
 \Big[
 {\cal F}_1^{ab}(p,p',\xi)\,\chi^{\dag}_{\kappa_a\mu_a}(\hp)\,\chi_{\kappa_b\mu_b}(\hpp)
\\ 
+{\cal F}_2^{ab}(p,p',\xi)\,\chi^{\dag}_{-\kappa_a\mu_a}(\hp)\,\chi_{-\kappa_b\mu_b}(\hpp)
\Big]\,,
\end{multline}
with explicit formulas for ${\cal F}_1$ and ${\cal F}_2$ given by Eqs.~(A5) and (A6)
of Ref.~\cite{yerokhin:99:sescr} and $p = |\bfp|$, $\hp = \bfp/p$,
and $\xi = \hp \cdot \hpp$. Performing angular integrations, we obtain
\begin{multline}\label{mom:3}
z_{\rm ver}^{(0)} = 
 {\cal P}_{ab}\, \frac{\alpha}{8\pi^{9/2}\rho^5}\, i^{l_a-l_b+1}\,
\\ \times 
    \int_0^{\infty}dp\,\int_0^{\infty}dp\pr\,\int_{-1}^{1}d\xi\,
    (p p\pr)^2\,e^{-\nicefrac{q^2}{\rho^2}}\,
\\ \times 
\bigg\{
    {\cal F}_1^{ab}(p,p',\xi)\,\Big[ p\pr P_{l_a}(\xi) - p P_{l_b}(\xi)\Big]
\\
  +  {\cal F}_2^{ab}(p,p',\xi)\,\Big[ p\pr P_{\overline{l}_a}(\xi) - p P_{\overline{l}_b}(\xi)\Big]
  \bigg\}\,,
\end{multline}
where $l_a = |\kappa_a+1/2|-1/2$, $\overline{l}_a = |\kappa_a-1/2|-1/2$ and $P_l(\xi)$ is the Legendre polynomial.
We note that $i^{l_a-l_b+1}$ is real since $l_a-l_b+1$ should be even, according to angular selection rules.

The remaining integrals in Eq.~(\ref{mom:3}) were performed numerically, after the change of variables
according to
\begin{multline}\label{mom:4}
    \int_0^{\infty}dp\, \int_0^{\infty}dp\pr\,\int_{-1}^{1}d\xi\,
    (p p\pr)^2 \, F(p,p',\xi)
\\
    = \int_0^{\infty}\!dx \int_0^x \!dy \int_y^x \!dq\, \frac{pp'q}{2}\, \Big[F(p,p',\xi)+F(p',p,\xi)\Big]\,,
\end{multline}
where $p = (x+y)/2$, $p' = (x-y)/2$, $q^2 = p^2+p^{\prime2}-2pp'\xi$.

\subsection{Frequency dependent correction}

In our calculations so far we used the low-frequency limit of the $E1$-transition
operator in the length gauge $D \propto r_z$, whose matrix elements reduce to the radial amplitude
$r_{ab}$ given by Eq.~\eqref{eq:rab}. 
The complete relativistic $E1$-transition operator contains in addition the frequency-dependent part 
\cite{Grant1974,Johnson07}, 
which is of order $O(2\pi r/\lambda)$ where $\lambda$ is the transition wave length. 
The frequency-dependent contribution is very small for neutral and slightly ionized
atoms, but become significant for heavy highly ionized ions \cite{Popov2017}.
In order to asses the uncertainty due to use of the low-frequency limit
of the transition operator, we now discuss the frequency-dependent correction
to the radial amplitude $r_{ab}$.

The radial amplitudes of the relativistic $E1$ transition operator in the length gauge were presented in Refs.~\cite{Grant1974,Johnson07} for the case of photon absorption:
\begin{multline}\label{eq:dab}
d_{ab} =  \frac{3}{k}\int_0^{\infty}\!r^2dr 
\bigg\{j_1(kr)\Big[ g_a(r)g_b(r)+f_a(r)f_b(r)\Big]
\\
+ j_2(kr)\Big[\frac{\kappa_a-\kappa_b+2}{2} g_a(r)f_b(r)
\\
+\frac{\kappa_a-\kappa_b-2}{2}f_a(r)g_b(r)\Big]
\bigg\}
\,,
\end{multline}
where $j_n(x)=\sqrt{\frac{\pi}{2x}}J_{n+1/2}(x)$ are spherical Bessel functions:
\begin{align}
    j_1(x) &= \frac{\sin x}{x^2} - \frac{\cos x}{x}
    = \frac{1}{3}x -\frac{1}{30}x^3 +\mathcal{O}(x^5)\,,
 \\
    j_2(x) &= \left(\frac{3}{x^2}-1\right)\frac{\sin x}{x} - \frac{3\cos x}{x^2}
\nonumber\\    
    &= \frac{1}{15}x^2 - \frac{1}{210}x^4 +\mathcal{O}(x^6)\,.
\end{align}
We immediately see that in the limit $kr\to 0$ only the first term in Eq.~(\ref{eq:dab}) survives and $d_{ab}$ reduces to $r_{ab}$, as it should.
Restricting ourselves to the next term of the expansion in $kr$,
we get the correction to the radial integral $d_{ab}-r_{ab}
\approx \delta d_{ab}$, where
\begin{multline}\label{eq:ddab}
\delta d_{ab} =  \int_0^{\infty}\!r^3dr 
\bigg\{-\frac{(kr)^2}{10} \Big[ g_a(r)g_b(r)+f_a(r)f_b(r)\Big]
\\
+ \frac{kr}{5}\Big[\frac{\kappa_a-\kappa_b+2}{2} g_a(r)f_b(r)
\\
+\frac{\kappa_a-\kappa_b-2}{2}f_a(r)g_b(r)\Big]
\bigg\}
\,.
\end{multline}
For the emission of the photon we need to substitute $k$ with $-k$ in the expressions \eqref{eq:dab} and \eqref{eq:ddab}. This keeps the first term in the integral \eqref{eq:ddab} unchanged, but changes sign of the second one. However, if we simultaneously interchange the initial and final states, $a \leftrightarrow b$, we get the same result: $\delta d_{ab}(k)=\delta d_{ba}(-k)$.

The frequency-dependent corrections are suppressed by the ratio of the size of the atom
to the transition wavelength. They are typically very small for most of experimentally studied
cases of valence transitions in neutral and slightly ionized atoms, as follows from a simple estimate.
For example, the $6s_{1/2}\to 6p_{1/2}$ transition in neutral Cs has frequency $\omega = 11178$~cm$^{-1}$. Taking into account the values of the rms radii of the $6s_{1/2}$ and $6p_{1/2}$ orbitals of $0.34$~nm and $0.46$ nm, respectively, the first term in Eq.\ \eqref{eq:ddab} is estimated to be about $\frac{(kr)^2}{10} r_{ab}\sim 8\cdot 10^{-7}r_{ab}$. The second term includes products of the upper and lower components, so it is about $\frac{(kr)\alpha}{5} r_{ab}\sim 4\cdot 10^{-6}r_{ab}$. 
The frequency-dependent correction for neutral cesium
is thus negligibly small even compared to the QED correction.

However, the situation changes drastically for
heavy highly ionized ions, because the transition energy grows typically as the square of the effective nuclear charge. Our explicit calculations presented in the next section show that 
for H-like cesium the frequency-dependent corrections enter already at the percent level.

\begin{table}
\caption{
Self-energy correction
to the decay rate of the $2p_{1/2}$ state, 
in terms of ${\cal R}_{\rm se}$ 
defined by Eq.~(\ref{res:3}).
\label{tab:1a}
}
\begin{ruledtabular}
\begin{tabular}{cw{2.8}w{2.3}w{2.5}}
 \multicolumn{1}{c}{$Z$} &
     \multicolumn{1}{c}{This work} &
          \multicolumn{1}{c}{Ref.\ \cite{SaPaChe04}} &
              \multicolumn{1}{c}{$\Za$-expansion}
                      \\
                         \hline\\[-5pt]
%
%
  2 &  -0.003\,55\,(4)  &        & -0.00343 \\
  3 &  -0.007\,00\,(3)  &        & -0.00670 \\
  4 &  -0.011\,23\,(2)  &        & -0.0106  \\
  5 &  -0.016\,13\,(3)  & -0.014 & -0.0151  \\
 10 &  -0.047\,62\,(5)  & -0.045 & -0.0409 \\
 20 &  -0.131\,25\,(2)  & -0.126 & -0.0860 \\
 30 &  -0.230\,11\,(2)  & -0.230 & -0.0917 \\
 40 &  -0.339\,52\,(2)  & -0.334 \\
 50 &  -0.459\,30\,(1)  & -0.449 \\
 60 &  -0.591\,32\,(1)  & -0.583 \\
 70 &  -0.738\,41\,(1)  & -0.738 \\
 80 &  -0.903\,81\,(1)  & -0.921 \\
 90 &  -1.090\,11\,(1)  & -1.144 \\
100 &  -1.296\,02\,(1)  & -1.426 \\
\end{tabular}
\end{ruledtabular}
\end{table}

\begin{table}
\caption{
Self-energy corrections
to the $E1$ amplitude of the $1s$-$2p_{1/2}$ transition of H-like ions, in terms
of $R_{\rm se}(\Za)$ defined by Eq.~(\ref{res:1}). $r_{ab}$
are the zeroth-order radial integrals defined by Eq.~(\ref{eq:rab}) in r.u. 
\label{tab:1}
}
\begin{ruledtabular}
\begin{tabular}{cw{1.4}w{1.5}w{1.5}w{1.7}w{1.7}}
 \multicolumn{1}{c}{$Z$} &
     \multicolumn{1}{c}{$|r_{ab}|$} &
     \multicolumn{1}{c}{po} &
         \multicolumn{1}{c}{vr, free} &
         \multicolumn{1}{c}{vr, many} &
             \multicolumn{1}{c}{se}
                 \\
                         \hline\\[-5pt]
 %
  2 &   88.3999 &  0.00553         &  -0.18863        &  0.18861\,(2)    &  0.00550\,(2)   \\
  3 &   58.9278 &  0.01095         &  -0.18777        &  0.18773\,(1)    &  0.01091\,(1)   \\
  4 &   44.1901 &  0.01764         &  -0.18674        &  0.18668\,(1)    &  0.01759\,(1)   \\
  5 &   35.3462 &  0.02539         &  -0.18558        &  0.18549\,(1)    &  0.02530\,(1)   \\
 10 &   17.6485 &  0.07578         &  -0.17876        &  0.17849\,(2)    &  0.07551\,(2)   \\
 20 &    8.7747 &  0.21153         &  -0.16408        &  0.16314\,(1)    &  0.21059\,(1)   \\
 30 &    5.7940 &  0.37233         &  -0.15086        &  0.14880\,(1)    &  0.37026\,(1)   \\
 40 &    4.2855 &  0.54808         &  -0.13929        &  0.13592\,(1)    &  0.54470\,(1)   \\
 50 &    3.3647 &  0.73564         &  -0.12840        &  0.12394\,(1)    &  0.73117\,(1)        \\
 55 &    3.0241 &  0.83391         &  -0.12283        &  0.11804\,(1)    &  0.82912\,(1)   \\
 60 &    2.7364 &  0.93556         &  -0.11692        &  0.11204\,(1)    &  0.93067\,(1)   \\
 70 &    2.2735 &  1.15148         &  -0.10336        &  0.09937         &  1.14749        \\
 80 &    1.9119 &  1.39117         &  -0.08601        &  0.08501         &  1.39017        \\
 90 &    1.6151 &  1.66955         &  -0.06269        &  0.06787         &  1.67474        \\
100 &    1.3600 &  2.01683         &  -0.03027        &  0.04659         &  2.03315        \\
\end{tabular}
\end{ruledtabular}
\end{table}

\begin{table*}
\caption{
Self-energy corrections to the $ns$-$n'p$ $E1$ matrix elements for H-like Cs ($Z = 55$) in terms of
$R_{\rm se}(\Za)$ defined by Eq.~(\ref{res:1}). $r_{ab}$ and $\delta d_{ab}$ are the zeroth-order radial integrals and frequency-dependent corrections defined by Eqs.\ \eqref{eq:rab} and \eqref{eq:ddab} respectively, in r.u.;
``qmod'' labels approximate results obtained with the QEDMOD package.
\label{tab:2n}
}
\begin{ruledtabular}
\begin{tabular}{ccw{3.4}w{4.2}w{3.6}w{4.6}w{4.8}w{3.4}w{2.1}w{2.1}}
 \multicolumn{1}{c}{$a$} &
     \multicolumn{1}{c}{$b$} &
     \multicolumn{1}{c}{$|r_{ab}|$} &
     \multicolumn{1}{c}{$10^3\frac{\delta d_{ab}(\omega)}{r_{ab}}$} &
         \multicolumn{1}{c}{po} &
             \multicolumn{1}{c}{vr} &
                 \multicolumn{1}{c}{se} &
                 \multicolumn{1}{c}{qmod} &
                 \multicolumn{1}{c}{$\frac{\mathrm{qmod-po}}{\mathrm{po}}$,\%} &
                 \multicolumn{1}{c}{$\frac{\mathrm{qmod-se}}{\mathrm{se}}$,\%} 
                      \\
                         \hline
 %
 %
 $1s$ & $2p_{1/2}$ &   3.0241 &  11.73 &  0.83391       &  -0.00480      &  0.82910\,(1)  &  0.866   &  3.8   &  4.4   \\
 $1s$ & $3p_{1/2}$ &   1.1647 &  20.73 &  0.54248       &  -0.04076\,(6) &  0.50172\,(6)  &  0.560   &  3.2   &  11.6  \\
 $1s$ & $4p_{1/2}$ &   0.6760 &  23.86 &  0.46690       &  -0.0573\,(3)  &  0.4096\,(3)   &  0.480   &  2.8   &  17.1  \\
 $1s$ & $5p_{1/2}$ &   0.4595 &  25.29 &  0.43673\,(5)  &  -0.0661\,(4)  &  0.3706\,(4)   &  0.446   &  2.2   &  20.4  \\
 $1s$ & $6p_{1/2}$ &   0.3450 &  26.07 &  0.4219\,(5)   &  -0.072\,(2)   &  0.350\,(2)    &  0.429   &  1.7   &  22.5  \\[2pt]
 $1s$ & $2p_{3/2}$ &   3.0161 & -36.04 &  0.81139       &  0.03740\,(1)  &  0.84879\,(1)  &  0.838   &  3.3   &  -1.2  \\
 $1s$ & $3p_{3/2}$ &   1.2341 & -36.05 &  0.59164       &  0.0089\,(1)   &  0.6006\,(1)   &  0.578   &  -2.4  &  -3.8  \\
 $1s$ & $4p_{3/2}$ &   0.7304 & -36.03 &  0.53484\,(1)  &  -0.0044\,(5)  &  0.5304\,(5)   &  0.500   &  -6.6  &  -5.8  \\
 $1s$ & $5p_{3/2}$ &   0.5010 & -36.03 &  0.51263\,(5)  &  -0.0118\,(5)  &  0.5008\,(5)   &  0.466   &  -9.1  &  -7.0  \\
 $1s$ & $6p_{3/2}$ &   0.3787 & -36.03 &  0.5020\,(6)   &  -0.016\,(1)   &  0.486\,(1)    &  0.449   &  -10.6 &  -7.7  \\[2pt]
 $2s$ & $2p_{1/2}$ &  12.0497 &   0.00 &  0.02819       &  0.02164       &  0.04983       &  0.054   &  92.7  &  9.0   \\
 $2s$ & $3p_{1/2}$ &   7.3677 &  -0.01 &  1.05192       &  0.01590\,(1)  &  1.06782\,(2)  &  1.078   &  2.5   &  1.0   \\
 $2s$ & $4p_{1/2}$ &   3.0501 &   2.98 &  0.66348       &  0.0061\,(1)   &  0.6696\,(1)   &  0.686   &  3.3   &  2.4   \\
 $2s$ & $5p_{1/2}$ &   1.8309 &   4.35 &  0.54219\,(5)  &  -0.001\,(2)   &  0.541\,(2)    &  0.558   &  3.0   &  3.3   \\
 $2s$ & $6p_{1/2}$ &   1.2931 &   5.09 &  0.4861\,(5)   &  -0.007\,(2)   &  0.479\,(2)    &  0.499   &  2.6   &  4.0   \\[2pt]
 $2s$ & $2p_{3/2}$ &  12.5733 &  -0.38 &  0.07397       &  0.02306\,(1)  &  0.09703\,(1)  &  0.100   &  35.0  &  2.9   \\
 $2s$ & $3p_{3/2}$ &   6.6065 &  -9.61 &  1.07766       &  0.0255\,(1)   &  1.1031\,(1)   &  1.100   &  2.0   &  -0.3  \\
 $2s$ & $4p_{3/2}$ &   2.8941 &  -9.80 &  0.73978\,(1)  &  0.0177\,(1)   &  0.7575\,(1)   &  0.752   &  1.6   &  -0.8  \\
 $2s$ & $5p_{3/2}$ &   1.7713 &  -9.88 &  0.63274\,(5)  &  0.012\,(3)    &  0.644\,(3)    &  0.637   &  0.7   &  -1.1  \\
 $2s$ & $6p_{3/2}$ &   1.2646 &  -9.92 &  0.5838\,(6)   &  0.008\,(2)    &  0.591\,(2)    &  0.583   &  -0.1  &  -1.4  \\[2pt]
 $3s$ & $2p_{1/2}$ &   2.2735 & -11.71 &  -1.89331\,(5) &  0.05923\,(7)  &  -1.83408\,(8) &  -1.855  &  -2.0  &  1.1   \\
 $3s$ & $3p_{1/2}$ &  30.1688 &   0.00 &  0.02006       &  0.01108\,(4)  &  0.03114\,(4)  &  0.030   &  47.3  &  -5.1  \\
 $3s$ & $4p_{1/2}$ &  13.2394 &  -0.81 &  1.16592\,(3)  &  0.0119\,(2)   &  1.1778\,(2)   &  1.185   &  1.6   &  0.6   \\
 $3s$ & $5p_{1/2}$ &   5.4498 &   0.55 &  0.75111\,(5)  &  0.009\,(2)    &  0.760\,(2)    &  0.768   &  2.2   &  1.0   \\
 $3s$ & $6p_{1/2}$ &   3.3201 &   1.29 &  0.6120\,(5)   &  0.006\,(1)    &  0.618\,(2)    &  0.625   &  2.2   &  1.2   \\[2pt]
 $3s$ & $2p_{3/2}$ &   2.9421 &  -1.35 &  -1.41360\,(4) &  0.04475\,(5)  &  -1.36884\,(7) &  -1.372  &  -3.0  &  0.2   \\
 $3s$ & $3p_{3/2}$ &  30.9663 &  -0.11 &  0.06368       &  0.0114\,(1)   &  0.0750\,(1)   &  0.075   &  18.5  &  0.6   \\
 $3s$ & $4p_{3/2}$ &  11.5084 &  -4.31 &  1.21398\,(3)  &  0.0162\,(2)   &  1.2302\,(2)   &  1.224   &  0.9   &  -0.5  \\
 $3s$ & $5p_{3/2}$ &   5.0114 &  -4.44 &  0.83882\,(5)  &  0.015\,(3)    &  0.854\,(3)    &  0.846   &  0.9   &  -0.9  \\
 $3s$ & $6p_{3/2}$ &   3.1179 &  -4.50 &  0.7107\,(6)   &  0.013\,(2)    &  0.723\,(2)    &  0.715   &  0.5   &  -1.2  \\[2pt]
 $4s$ & $2p_{1/2}$ &   0.9285 & -12.73 &  -1.61639\,(1) &  0.0777\,(3)   &  -1.5387\,(3)  &  -1.565  &  -3.2  &  1.7   \\
 $4s$ & $3p_{1/2}$ &   5.8984 &  -4.90 &  -1.75260\,(2) &  0.0260\,(4)   &  -1.7266\,(4)  &  -1.726  &  -1.5  &  0.0   \\
 $4s$ & $4p_{1/2}$ &  55.7204 &   0.00 &  0.01608       &  0.00658\,(4)  &  0.02266\,(4)  &  0.020   &  25.9  &  -10.7 \\
 $4s$ & $5p_{1/2}$ &  20.7063 &  -0.74 &  1.23126\,(3)  &  0.008\,(3)    &  1.239\,(3)    &  1.244   &  1.0   &  0.4   \\
 $4s$ & $6p_{1/2}$ &   8.4931 &  -0.01 &  0.8089\,(4)   &  0.0076\,(9)   &  0.817\,(1)    &  0.822   &  1.6   &  0.6   \\[2pt]
 $4s$ & $2p_{3/2}$ &   1.1473 &   1.01 &  -1.15431\,(1) &  0.0581\,(3)   &  -1.0962\,(3)  &  -1.100  &  -4.7  &  0.3   \\
 $4s$ & $3p_{3/2}$ &   7.4789 &  -1.15 &  -1.33755\,(2) &  0.0204\,(4)   &  -1.3172\,(4)  &  -1.310  &  -2.0  &  -0.5  \\
 $4s$ & $4p_{3/2}$ &  56.7637 &  -0.05 &  0.05716       &  0.00668\,(9)  &  0.06384\,(9)  &  0.064   &  11.3  &  -0.3  \\
 $4s$ & $5p_{3/2}$ &  17.7224 &  -2.44 &  1.29323\,(4)  &  0.011\,(2)    &  1.304\,(2)    &  1.291   &  -0.2  &  -1.0  \\
 $4s$ & $6p_{3/2}$ &   7.6985 &  -2.52 &  0.9047\,(5)   &  0.011\,(1)    &  0.916\,(1)    &  0.905   &  0.0   &  -1.2  \\[2pt]
 $5s$ & $2p_{1/2}$ &   0.5542 & -13.20 &  -1.53456\,(5) &  0.083\,(2)    &  -1.451\,(2)   &  -1.481  &  -3.5  &  2.1   \\
 $5s$ & $3p_{1/2}$ &   2.3455 &  -5.41 &  -1.44462\,(4) &  0.037\,(4)    &  -1.408\,(4)   &  -1.407  &  -2.6  &  0.0   \\
 $5s$ & $4p_{1/2}$ &  11.1399 &  -2.63 &  -1.67961\,(4) &  0.014\,(6)    &  -1.665\,(6)   &  -1.665  &  -0.8  &  0.0   \\
 $5s$ & $5p_{1/2}$ &  88.7317 &   0.00 &  0.01350\,(1)  &  0.00\,(2)     &  0.02\,(2)     &  0.016   &  15.2  &  -9.7  \\
 $5s$ & $6p_{1/2}$ &  30.0484 &  -0.60 &  1.2729\,(3)   &  0.007\,(4)    &  1.280\,(4)    &  1.282   &  0.7   &  0.2   \\[2pt]
 $5s$ & $2p_{3/2}$ &   0.6751 &   2.10 &  -1.07852\,(5) &  0.062\,(2)    &  -1.017\,(2)   &  -1.022  &  -5.2  &  0.5   \\
 $5s$ & $3p_{3/2}$ &   2.8397 &  -0.08 &  -1.04889\,(3) &  0.028\,(3)    &  -1.021\,(3)   &  -1.013  &  -3.4  &  -0.7  \\
 $5s$ & $4p_{3/2}$ &  13.9450 &  -0.84 &  -1.29713\,(4) &  0.012\,(3)    &  -1.285\,(3)   &  -1.271  &  -2.0  &  -1.1  \\
 $5s$ & $5p_{3/2}$ &  90.0012 &  -0.03 &  0.05275\,(1)  &  0.005\,(4)    &  0.058\,(4)    &  0.057   &  7.5   &  -2.5  \\
 $5s$ & $6p_{3/2}$ &  25.5038 &  -1.56 &  1.3443\,(3)   &  0.010\,(6)    &  1.354\,(6)    &  1.333   &  -0.8  &  -1.5  \\[2pt]
 $6s$ & $2p_{1/2}$ &   0.3902 & -13.44 &  -1.4959\,(7)  &  0.085\,(2)    &  -1.411\,(2)   &  -1.443  &  -3.6  &  2.3   \\
 $6s$ & $3p_{1/2}$ &   1.4054 &  -5.68 &  -1.3534\,(6)  &  0.041\,(2)    &  -1.313\,(2)   &  -1.312  &  -3.0  &  0.0   \\
 $6s$ & $4p_{1/2}$ &   4.3917 &  -2.91 &  -1.3525\,(5)  &  0.020\,(2)    &  -1.332\,(2)   &  -1.331  &  -1.6  &  -0.1  \\
 $6s$ & $5p_{1/2}$ &  18.1437 &  -1.62 &  -1.6352\,(3)  &  0.011\,(6)    &  -1.624\,(6)   &  -1.627  &  -0.5  &  0.2   \\
 $6s$ & $6p_{1/2}$ & 127.2007 &   0.00 &  0.0117\,(3)   &  0.003\,(3)    &  0.015\,(3)    &  0.013   &  8.8   &  -15.1 \\[2pt]
 $6s$ & $2p_{3/2}$ &   0.4723 &   2.69 &  -1.0426\,(7)  &  0.0620\,(7)   &  -0.981\,(1)   &  -0.987  &  -5.4  &  0.6   \\
 $6s$ & $3p_{3/2}$ &   1.6771 &   0.50 &  -0.9647\,(6)  &  0.030\,(2)    &  -0.934\,(2)   &  -0.926  &  -4.0  &  -0.9  \\
 $6s$ & $4p_{3/2}$ &   5.2457 &  -0.25 &  -0.9907\,(5)  &  0.016\,(1)    &  -0.975\,(2)   &  -0.961  &  -3.0  &  -1.5  \\
 $6s$ & $5p_{3/2}$ &  22.5169 &  -0.62 &  -1.2720\,(2)  &  0.009\,(3)    &  -1.263\,(3)   &  -1.245  &  -2.1  &  -1.4  \\
 $6s$ & $6p_{3/2}$ & 128.6106 &  -0.01 &  0.0496\,(3)   &  0.004\,(5)    &  0.054\,(5)    &  0.052   &  5.2   &  -2.8  \\
\end{tabular}
\end{ruledtabular}
\end{table*}

\begin{table*}
\caption{
Self-energy corrections to $np$-$n'd$ $E1$ matrix elements in H-like Cs.
Notations and units are as in Table~\ref{tab:2n}.
\label{tab:3}
}
\begin{ruledtabular}
\begin{tabular}{ccw{3.4}w{4.2}w{3.6}w{4.6}w{4.8}w{3.4}w{2.1}w{2.1}}
 \multicolumn{1}{c}{$a$} &
     \multicolumn{1}{c}{$b$} &
     \multicolumn{1}{c}{$|r_{ab}|$} &
     \multicolumn{1}{c}{$10^3\frac{\delta d_{ab}(\omega)}{r_{ab}}$} &
         \multicolumn{1}{c}{po} &
             \multicolumn{1}{c}{vr} &
                 \multicolumn{1}{c}{se} &
                 \multicolumn{1}{c}{qmod} &
                 \multicolumn{1}{c}{$\frac{\mathrm{qmod-po}}{\mathrm{po}}$,\%} &
                 \multicolumn{1}{c}{$\frac{\mathrm{qmod-se}}{\mathrm{se}}$,\%} 
                      \\
                         \hline
 %
 %
 $2p_{1/2}$ & $3d_{3/2}$ &  10.8136 & -3.28 &  0.01139       &  0.00526\,(3)  &  0.01665\,(3)  &  0.024   &  108.6  &  42.7     \\
 $2p_{1/2}$ & $4d_{3/2}$ &   4.0780 & -1.60 &  0.00961\,(1)  &  -0.0105\,(3)  &  -0.0009\,(3)  &  0.009   &  -9.2   &  -1075.9  \\
 $2p_{1/2}$ & $5d_{3/2}$ &   2.3589 & -0.82 &  0.01274\,(2)  &  -0.0199\,(1)  &  -0.0071\,(1)  &  0.002   &  -86.7  &  -123.8   \\[3pt]
 $2p_{3/2}$ & $3d_{3/2}$ &  11.5701 &  5.53 &  0.05593       &  -0.00154\,(7) &  0.05439\,(7)  &  0.069   &  24.2   &  27.8     \\
 $2p_{3/2}$ & $4d_{3/2}$ &   4.0892 & 10.41 &  0.00906\,(1)  &  -0.0211\,(3)  &  -0.0120\,(3)  &  0.007   &  -24.4  &  -157.0   \\
 $2p_{3/2}$ & $5d_{3/2}$ &   2.3118 & 12.68 &  -0.00309\,(2) &  -0.03228\,(4) &  -0.03536\,(5) &  -0.017  &  454.9  &  -51.5    \\[3pt]
 $3p_{1/2}$ & $3d_{3/2}$ &  24.7424 & -0.03 &  -0.01610      &  0.01015\,(8)  &  -0.00595\,(8) &  -0.009  &  -41.2  &  59.1     \\
 $3p_{1/2}$ & $4d_{3/2}$ &  16.6721 & -2.06 &  0.03348       &  0.0081\,(1)   &  0.0416\,(1)   &  0.048   &  42.1   &  14.4     \\
 $3p_{1/2}$ & $5d_{3/2}$ &   6.9098 & -1.28 &  0.01781\,(2)  &  0.0016\,(2)   &  0.0194\,(2)   &  0.024   &  32.8   &  21.8     \\[3pt]
 $3p_{3/2}$ & $3d_{3/2}$ &  24.5288 &  0.00 &  -0.02471      &  0.01077\,(5)  &  -0.01393\,(5) &  -0.017  &  -31.1  &  22.2     \\
 $3p_{3/2}$ & $4d_{3/2}$ &  18.6295 &  1.17 &  0.09649\,(1)  &  0.0052\,(1)   &  0.1017\,(1)   &  0.113   &  17.4   &  11.5     \\
 $3p_{3/2}$ & $5d_{3/2}$ &   7.2515 &  3.41 &  0.04023\,(2)  &  -0.0027\,(2)  &  0.0375\,(2)   &  0.048   &  19.7   &  28.4     \\[3pt]
 $4p_{1/2}$ & $3d_{3/2}$ &   4.1851 & -2.81 &  -0.12454\,(1) &  0.0258\,(4)   &  -0.0988\,(4)  &  -0.095  &  -24.1  &  -4.2     \\
 $4p_{1/2}$ & $4d_{3/2}$ &  51.1988 & -0.01 &  -0.00999      &  0.00610\,(2)  &  -0.00389\,(2) &  -0.007  &  -28.4  &  83.8     \\
 $4p_{1/2}$ & $5d_{3/2}$ &  23.8664 & -1.39 &  0.04863\,(2)  &  0.0068\,(2)   &  0.0555\,(2)   &  0.060   &  23.6   &  8.4      \\[3pt]
 $4p_{3/2}$ & $3d_{3/2}$ &   3.2484 & -6.70 &  -0.32191\,(3) &  0.0337\,(5)   &  -0.2882\,(5)  &  -0.302  &  -6.1   &  4.9      \\
 $4p_{3/2}$ & $4d_{3/2}$ &  50.9107 &  0.00 &  -0.01661      &  0.00641\,(6)  &  -0.01020\,(6) &  -0.013  &  -20.0  &  30.2     \\
 $4p_{3/2}$ & $5d_{3/2}$ &  27.2582 &  0.18 &  0.11999\,(2)  &  0.00508\,(3)  &  0.12507\,(4)  &  0.140   &  16.5   &  11.8     \\[3pt]
 $5p_{1/2}$ & $3d_{3/2}$ &   1.4993 & -2.54 &  -0.12417\,(3) &  0.0394\,(5)   &  -0.0848\,(5)  &  -0.079  &  -36.5  &  -7.0     \\
 $5p_{1/2}$ & $4d_{3/2}$ &   9.5709 & -1.62 &  -0.10980\,(4) &  0.01393\,(9)  &  -0.0959\,(1)  &  -0.097  &  -11.7  &  1.1      \\
 $5p_{1/2}$ & $5d_{3/2}$ &  84.7700 & -0.01 &  -0.00673\,(1) &  0.00408\,(5)  &  -0.00265\,(5) &  -0.005  &  -20.8  &  101.5    \\[3pt]
 $5p_{3/2}$ & $3d_{3/2}$ &   1.2081 & -8.09 &  -0.30312\,(4) &  0.051\,(1)    &  -0.252\,(1)   &  -0.272  &  -10.3  &  7.9      \\
 $5p_{3/2}$ & $4d_{3/2}$ &   7.5474 & -3.48 &  -0.28786\,(7) &  0.0177\,(2)   &  -0.2702\,(2)  &  -0.297  &  3.3    &  10.1     \\
 $5p_{3/2}$ & $5d_{3/2}$ &  84.4356 &  0.00 &  -0.01228\,(1) &  0.0042\,(2)   &  -0.0081\,(2)  &  -0.011  &  -11.4  &  34.8     \\
\end{tabular}
\end{ruledtabular}
\end{table*}

\section{Results and discussion}

It is convenient to represent numerical results for the self-energy correction to the $E1$ transition amplitude
in terms of the multiplicative function $R_{\rm se}(\Za)$ defined as
\begin{align}\label{res:1}
\delta z_{\rm se} = z_{ab}\, \frac{\alpha}{\pi}\,R_{\rm se}(\Za)\,.
\end{align}

We start with comparing our numerical results with the numerical values obtained for the $2p_{1/2}$-$1s$ transition by Sapirstein, Pachucki and Cheng \cite{SaPaChe04} and with the leading terms of the $\Za$ expansion derived in that work. In that work results were obtained in the form of corrections to the $2p_{1/2}$-state decay rate, rather than amplitudes, so we need to connect the transition amplitude $z_{ab}$ to the decay rate $\Gamma_{ab}$. This is achieved by observing that
\begin{align}
\frac{\delta \Gamma_{ab}}{\Gamma_{ab}} = 3\,\frac{\delta E_{ab}}{E_{ab}} + 2\,\frac{\delta z_{ab}}{z_{ab}}\,,
\end{align}
where $E_{ab}$ and $\delta E_{ab}$ is the zero-order energy difference between the states $a$ and $b$ and its correction, respectively. The self-energy correction to $\Gamma_{ab}$ can then
be represented in a form analogous to Eq.~(\ref{res:1}),
\begin{align}\label{res:3}
\delta \Gamma_{\rm se} = \Gamma_{ab}\, \frac{\alpha}{\pi}\,{\cal R}_{\rm se}(\Za)\,.
\end{align}
To the leading order in $\Za$, the self-energy correction to the $2p_{1/2}$ decay rate is \cite{SaPaChe04,ivanov:06:pla}
\begin{align}
{\cal R}_{\rm se}(\Za)
= (\Za)^2
\Big\{ \Big[ \frac{8}{3}\ln \frac43-\frac{61}{18}\Big]
  \ln (\Za)^{-2} + 6.05168
 \Big\}.
\end{align}

Results of our numerical calculations of the self-energy correction to the decay rate of the $2p_{1/2}$ state are presented in Table~\ref{tab:1a}, in comparison with
predictions of the $\Za$ expansion and all-order numerical results obtained in Ref.~\cite{SaPaChe04}. We observe that both numerical calculations converge to the $\Za$-expansion predictions in the low-$Z$ limit but significantly deviate
from them for medium- and high-$Z$ values. The two numerical calculations are in reasonable agreement with each other.
We observe small deviations increasing with the increase of $Z$, probably due to different nuclear models used
in the calculations. (Ref.~\cite{SaPaChe04} used a finite-size nucleus, whereas we use the point nuclear model.)
Small deviations observed for the lowest $Z$ are probably due to numerical issues. It should be mentioned that Ref.~\cite{SaPaChe04} used a completely different approach to the calculation of the decay rates, via the imaginary part of the two-loop self-energy. The observed agreement of the two calculations is also a check of consistency of the two different approaches.

Table~\ref{tab:1} presents results of our numerical calculations of the self-energy 
correction to the $E1$ matrix elements of the $1s$-$2p_{1/2}$ transition along the hydrogen isoelectronic sequence. 
We observe large cancellations between the free-electron and many-potential parts of the vr correction, thus corroborating the findings of Ref.~\cite{SaPaChe04}, which reported large numerical cancellations at intermediate stages of calculations.
These cancellations are particularly pronounced in the low-$Z$ region. For instance, at $Z = 2$, the first four digits cancel out.
This highlights the need for meticulous control of numerical accuracy throughout the intermediate stages of these calculations.
We also observe that the vr contribution is much smaller than the po correction
in the whole region of nuclear charges, accounting for typically about 1\% of the total self-energy correction.

\begin{figure*}[htb!]
   \includegraphics[width=0.9\columnwidth]{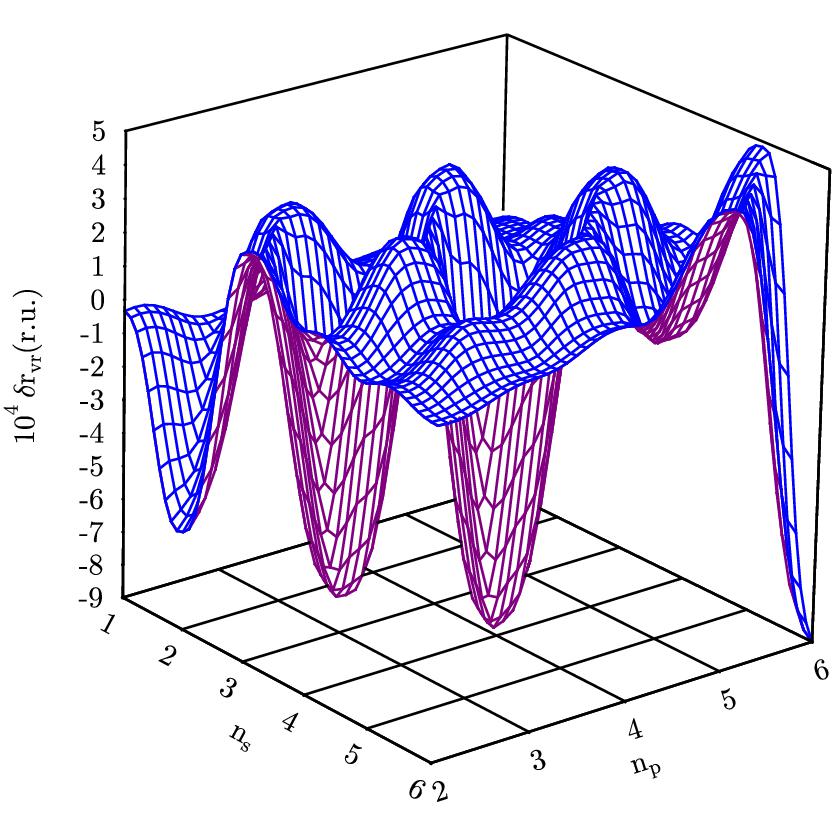}
    \hfill
    \includegraphics[width=0.9\columnwidth]{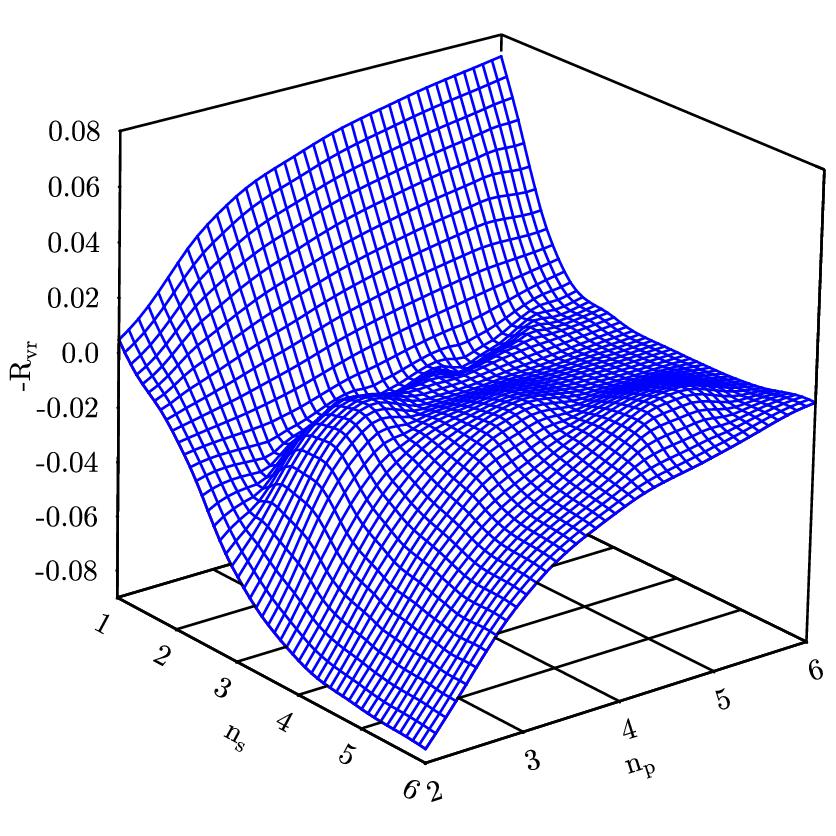}
    \caption{Vertex-reducible (vr) QED correction to the $n_s s_{1/2}\to n_p p_{1/2}$ transitions in H-like Cs ion. Left panel: corrections to the radial integrals $r_{ab}$; right panel: corrections to the scaled function $R_{\rm se}$ defined by Eq.~(\ref{res:1}), multiplied by (-1) for better vision.}
    \label{fig:R_vr}
\end{figure*}

We now turn to the detailed analysis of the $E1$ transition matrix elements between different states
of a selected H-like ion. We have chosen the reference ion to be cesium ($Z = 55$), since it is well studied experimentally
(see, e.\ g.\ Refs.\ \cite{TDGQ19,TCBD19,TDTJ19,QJDT24})
and its nuclear charge is large enough for QED effects to be of 
experimental interest.
Table~\ref{tab:2n} presents
our numerical results for the $ns$-$n'p$ matrix elements, whereas
Table~\ref{tab:3} summarizes our results obtained for the $np$-$n'd$ transitions. 

Tables \ref{tab:2n} and \ref{tab:3} include values for the relative frequency-dependent correction $\frac{\delta d_{ab}(\omega)}{r_{ab}}$, multiplied by $10^3$. They are calculated using Eq.\ \eqref{eq:ddab} and
taking the transition energy as the difference
of the Dirac energies of the initial and the final state.
We observe that the frequency-dependent correction is typically at a percent (sometime,
a few percent) level and thus should
be taken into account for H-like ions.
By contrast, it becomes negligibly small
for neutral and slightly ionized ions.
For example, if we recalculate this correction
with the same hydrogenic wave functions but with
frequency $\omega = 11178$~cm$^{-1}$ corresponding the $6s_{1/2} \to 6p_{1/2}$ transition in neutral Cs, we obtain the 
relative contribution on the $10^{-6}$ level,
which is much smaller than the QED corrections.

Table~\ref{tab:2n} and \ref{tab:3} presents
our numerical results for the self-energy correction for the $ns$-$n'p$ and $np$-$n'd$ transitions. 
We observe that the dependence of the self-energy corrections of the $(n,n')$ matrix elements
on $n$ and $n'$ turns out to be rather complicated. 
In particular, for the $(n,n)$ matrix elements the self-energy corrections are abnormally suppressed, even after taking into account that the zeroth-order radial 
integrals $r_{ab}$ reach their maximal values in this case. 
Furthermore,  for $n<n'$ and $n>n'$ the corrections are typically of the opposite sign. We conclude that the self-energy correction to the $E1$ matrix elements does not have a well-defined sign and can
both increase and decrease the absolute values of the transition amplitude. 

In the last three columns of Tables~\ref{tab:2n} and \ref{tab:3} we compare our ab initio values with approximate self-energy results obtained with help of QEDMOD package (with the vacuum-polarization part switched off). We recall that the QEDMOD results should approximately reproduce the po part of the ab inito se correction, whereas the vr part is omitted in the approximate treatment. 

We see that for the $ns$-$n'p$ matrix elements QEDMOD
calculations reproduce the po values very well in most cases. 
The exception is the diagonal $(n,n)$ matrix elements, where
the deviation may reach 90\% (for the $(2s,2p_{1/2})$ transition). 
In this case the po contribution is abnormally small, and
the absolute error turns out to be not very significant. 
The vr part of the se correction is also relatively small 
for the $ns$-$n'p$ matrix elements. As a result, we may conclude
that the 
QEDMOD treatment, despite ignoring the vr contribution, still
reproduces the ab inito se correction reasonably well,
the typical accuracy being within 10\% in most cases.

For the  $np$-$n'd$ matrix elements, however, 
the situation is
different. First,
the po part is significantly smaller in magnitude than for the $ns$-$n'p$ matrix elements, leading to significantly smaller se corrections. Next, 
the po and vr parts  are of the same order.
Furthermore, the QEDMOD calculations do not reproduce well
even the po part of the se correction. We conclude that
for the $np$-$n'd$ matrix elements, the QEDMOD treatment
provides only the order of magnitude of the se contribution,
but is not capable of yielding a quantitative approximation.

It is important to note that the situation in
many-electron atoms can differ significantly from that of hydrogen-like ions. In particular, many-electron atoms often exhibit substantial mixing of single-electron configurations due to electron correlation effects. For the $p$-$d$ transitions, where QED effects on radial integrals are relatively small, a more significant QED contribution may arise from the configuration mixing.
For example, this was the case in our recent study of the ten-electron ions Fe$^{16+}$ and Ni$^{18+}$ \cite{KoYeKa24}, where the dominant QED correction to the $2p$-$3d$ transitions was found to originate from electronic correlations. It was demonstrated that such corrections can be effectively accounted for by incorporating the QEDMOD potential in many-electron calculations.

The matrix elements of the QED corrections to the transition amplitudes in the hydrogenic basis, obtained in this
work, could, in principle, be used to construct a model operator analogous to QEDMOD but specifically tailored for 
the $E1$ transition amplitude. As noted earlier, the po part of these QED corrections can be sufficiently accurately accounted for by using the QEDMOD potential. 
Therefore, an additional model operator would only need to account for the vr correction.
To evaluate the feasibility of such a project, Fig.~\ref{fig:R_vr} presents the vr correction for the $n_s s_{1/2} \to n_p p_{1/2}$ transitions in hydrogen-like cesium as a function of the quantum numbers $n_s$ and $n_p$.
The left panel displays the vr correction to the radial integrals $r_{ab}$, while the right panel shows these corrections to the scaled function $R_{\rm se}$. 
The corrections to the radial integrals exhibit highly complex and irregular behavior. 
The dependence of the
scaled function is more regular but nevertheless remains quite complex. 
Notably, the sign of the correction differs between cases where $n_s < n_p$ and $n_s > n_p$.
This complex behavior might indicate that the vr correction originates not only from small
distances, as is typical for QED effects, but also from large distances. 
The intricate dependence of the vr correction on the principal quantum numbers of 
the transition states significantly complicates the construction of an effective operator.

\section{Summary}

We performed ab initio calculations of the self-energy
correction to the $E1$ transition amplitudes
in H-like ions, to all orders in the binding nuclear strength 
parameter $\Za$. 
Our results for the $1s$-$2p_{1/2}$ transition amplitude were
converted to the correction to the decay rate of the $2p_{1/2}$ state and compared
with previous all-order and $\Za$-expansion calculations \cite{SaPaChe04}.   
Good agreement was found, which yields an important check of consistency of different methods.
Our calculation confirms the previous conclusion of Ref.~\cite{SaChe05}
that the vertex and reducible parts of the self-energy 
correction are small, 
their contribution being on the level of
1\% for the $1s$-$2p_{1/2}$ matrix element.

A detailed study of self-energy correction for a large number of $ns$-$n'p$ and $np$-$n'd$
transitions was carried out for H-like cesium. 
We found a complicated dependence of the self-energy corrections of the $(n,n')$ matrix elements
on $n$ and $n'$. 
We conclude that for the $ns$-$n'p$ matrix elements, the perturbed-orbital contribution is dominant and the
vertex and reducible corrections are relatively small, so that
the total self-energy effect can usually be well 
reproduced by approximate calculations with the QEDMOD 
model potential. 
By contrast, for the $np$-$n'd$ matrix elements
we find significantly smaller self-energy corrections,
which are not  well reproduced by the QEDMOD 
model potential. 

The comparison of our ab initio results with approximate treatment based on the QEDMOD model potential allows 
one to estimate uncertainties associated with usage of
the QEDMOD potential in many-electron calculations
of the $E1$ transition amplitudes in multi-electron
atoms \cite{KCO22,SKBS24,Fairhall2023,Roberts2023,TaDe23}.

\begin{acknowledgments}
We are grateful to Andrey Bondarev and to Ilya Tupitsyn for the help with calculations of the frequency-dependent corrections to the radial integrals. This work was supported by the Russian Science Foundation grant \#23-22-00079. 
\end{acknowledgments}

\bibliography{./b-spline}

\end{document}